\documentclass[12pt,aps,prd,superscriptaddress,showpacs,longbibliography,floatfix,nofootinbib]{revtex4-2}

\usepackage[utf8]{inputenc}
\usepackage{slashed}
\pdfoutput=1

\usepackage{color}
\usepackage{graphicx}   
\usepackage{bm}
\usepackage{amsmath}
\usepackage{amsfonts}
\usepackage{hyperref}

\def\be{\begin{equation}}
\def\ee{\end{equation}}
\def\ba{\begin{eqnarray}}
\def\ea{\end{eqnarray}}

\begin{document}

\title{Testing Scalar Field Self-Dualities in d=2 using a Variational Method}

\author{Paul Romatschke}
\affiliation{Institut für Theoretische Physik, TU Wien, Wiedner Hauptstraße 8-10, 1040 Wien, Austria}
\affiliation{Department of Physics, University of Colorado, Boulder, Colorado 80309, USA}
\affiliation{Center for Theory of Quantum Matter,
University of Colorado, Boulder, Colorado 80309, USA}

\author{Ulrike Romatschke}
\affiliation{GeoSphere Austria, Climate Monitoring, Vienna, Austria}
\affiliation{NSF National Center for Atmospheric Research, Earth Observing Laboratory, Boulder, CO, USA}

\begin{abstract}
Recently, self-dualities based on saddle-point expansions have been proposed as a means to obtain qualitative non-perturbative information in scalar field theories. In this work, we test this proposition quantitatively by studying the phase transition for critical scalar $\phi^4$ field theory in 1+1 dimensions using a variational method. We find that saddle-point methods obtain quantitative agreement for the free energy, but differ on the order of 25 percent for the peak location of the correlation length.
\end{abstract}

\maketitle

\section{Introduction}

Interacting scalar field theory in 1+1 dimensions possesses a critical point that can be accessed by tuning bare parameters in a lattice theory \cite{Loinaz:1997az}. This makes the theory an interesting case for testing other non-perturbative methods on their quantitative performance in a non-trivial application.

In this work, our aim is to test the analytic saddle-point expansions introduced in Ref.~\cite{Romatschke:2026tam} on both the qualitative and quantitative level. Because the analytic expansions are directly applicable to more complicated cases, such as the study of critical phenomena in higher dimensions, knowledge of the quantitative performance of the method in two dimensions provides useful information on the reliability of the method. The analytic saddle-point expansions exhibit self-duality under a flip of the sign of the interaction parameter \cite{Romatschke:2026tam}, which historically has been the namesake for the approach \cite{Chang:1976ek,Serone:2018gjo}. 

To provide detailed non-perturbative information for the lattice theory, we introduce a variational method that allows diagonalization of the Hamiltonian for small transverse lattice sizes. The variational method we choose is straightforward to implement, robust, and flexible in regards to generalization of the system under study. However, our numerical implementation of the method is not optimized for extracting continuum quantities, so comparisons will be limited to small lattice sizes.

\section{Self-Dualities for a Finite-Volume Lattice System}

We want to study self-dualities for scalar field theory in d=2 defined by the Euclidean action 
\be
\label{mainaction}
S=\int d^2x \left[\frac{1}{2}\partial_\mu \phi \partial_\mu \phi+\frac{m_B^2 \phi^2}{2}+\lambda \phi^4\right]\,,
\ee
where $m_B^2$ is the bare (unrenormalized) mass parameter for the theory. This theory has been studied extensively in the literature and it is known to undergo a phase transition from a symmetric to a broken phase  at a critical  value of \cite{Schaich:2009jk}
\be
\label{gvalue}
g_{c}\equiv \frac{ \lambda}{m_{\rm crit}^2}\simeq 2.7^{+0.025}_{-0.0125}\,, 
\ee
where $m_{\rm crit}^2$ is the critical renormalized mass parameter defined by
\be
m_{\rm crit}^2=m_{B,{\rm crit}}^2+12 \lambda G_{\rm free}(m_{\rm crit})\,,
\label{gap}
\ee
where $G_{\rm free}(m)$ is the  two-point function for a free boson of mass $m$,
\be
\label{gfree}
G_{\rm free}(m)=\int \frac{d^2k}{(2\pi)^2}\frac{1}{k^2+m^2}\,.
\ee
One should point out that $g_c$ is found by tuning the only dimensionless parameter in the action, namely $\frac{m_B^2}{\lambda}$, and then finding $m_{\rm crit}^2$ numerically from (\ref{gap}) before reporting a value of $g_c$. We study this theory on a lattice by discretizing the action on $D$ points along one of the ``spatial'' Euclidean directions with length $L$, such that
\be
\label{latticeaction}
S_{(D)}=a \int d\tau \sum_{k=0}^{D-1} \left[\frac{\dot{\phi}^2(\tau,x_k)}{2}+\frac{\left(\phi(\tau,x_{k+1})-\phi(\tau,x_k)\right)^2}{2a^2}+\frac{m_B^2 \phi^2(\tau,x_k)}{2}+\lambda \phi^4\left(\tau,x_k\right)\right]\,,
\ee
where $a=\frac{L}{D}$ is the lattice spacing, $x_k=-\frac{L}{2}+k a$ with $k\in \mathbb{N}$ is the location on the discretized Euclidean direction, and $\dot{\phi}=\frac{d\phi}{d \tau}$ is the (continuous) derivative along the ``time'' Euclidean direction and $\lim_{D\rightarrow \infty} S_{(D)}=S$.

For the saddle-point expansions introduced in Ref.~\cite{Romatschke:2026tam}, it is sufficient to characterize the pressure function of the free theory of the lattice theory, defined by
\be
p_{\rm free}(m_B)=\left.\frac{\ln Z}{\beta L}\right|_{\lambda=0}\,,
\ee
where $Z=\int {\cal D}\phi e^{-S_{(D)}}$ is the partition function of the theory and $\beta$ is the length of the Euclidean time direction $\beta=\int d\tau$. Using
\be
\phi(\tau,x)=\frac{1}{L}\sum_n \int \frac{d\omega}{2\pi} e^{i\omega \tau+i k_n x}\phi(\omega, k_n)\,,
\ee
with $k_n=\frac{2\pi}{L}n$, $n\in \mathbb{N}$ the bosonic Matsubara frequencies, one has
\be
S_{(D)}=\frac{1}{2L}\sum_n \int \frac{d\omega}{2\pi}\left(\omega^2+m_B^2+\Omega_n^2(D)\right)\left|\phi\right|^2\,,
\ee
where
\be
\Omega_n^2(D)=\frac{2}{a^2}\left(1-\cos\left(\frac{2\pi n}{D}\right)\right)\,.
\ee
For a lattice with $D$ points there are exactly $D$ Matsubara frequencies, e.g. for $D=3$ we have $n=\left\{-1,0,1\right\}$.
As a consequence, the free theory pressure becomes \cite{Laine:2016hma}
\be
p_{\rm free}(m_B)=-\frac{1}{2L}\sum_n \int \frac{d\omega}{2\pi}\ln\left[\omega^2+m_B^2+\Omega_n^2(D)\right]=-\frac{1}{2L}\sum_n \sqrt{m_B^2+\Omega_n^2(D)}\,.
\ee

Using this expression, we follow the same steps as in Ref.~\cite{Romatschke:2026tam} to find the pressure for the \textbf{interacting theory} $\lambda\neq 0$ in the symmetric and broken phase in the R1-level resummation scheme \cite{Romatschke:2019rjk} for a finite-volume lattice system:
\ba
\label{saddlepoints}
p(M)=p_{\rm free}(M)+\frac{(M^2-m_B^2)^2}{48\lambda}\,,\quad M^2=m_B^2+12 \lambda G_{\rm free}(M)\\
\tilde p(\tilde M)=p_{\rm free}(\tilde M)-\frac{\tilde M^4}{96 \lambda}-\frac{\tilde M^2 m_B^2}{24 \lambda}+\frac{m_B^4}{48\lambda}\,,\quad -\frac{\tilde M^2}{2}=m_B^2+12 \lambda G_{\rm free}(\tilde M)\,.
\ea
Here $G_{\rm free}(m)$ is the lattice-version of (\ref{gfree}), given by \cite{Laine:2016hma}
\be
\label{gfreedef}
G_{\rm free}(m)=-2 \frac{\partial p_{\rm free}(m)}{\partial m^2}=\frac{1}{2L}\sum_n \frac{1}{\sqrt{m^2+\Omega_n^2(D)}}\,.
\ee

Note that both the symmetric saddle pressure $p(M)$ and the broken phase saddle pressure $\tilde p(\tilde M)$ are readily evaluated for fixed parameters $D, \lambda, m_B^2$. For the broken-phase saddle, solutions to the saddle-point condition $\tilde M$ are not always real-valued for all parameters. In addition, when $\tilde M$ is real-valued, more than one solution exists. Comparing the difference in pressures between solutions, we follow Ref.~\cite{Romatschke:2026tam} to select the solution that has the lowest free energy (highest pressure) for a given set of parameters to represent the thermodynamically preferred phase, even though we are able to track the pressure for thermodynamically disfavored phases as well.

\section{Variational Method}

\subsection{Warm-up: quantum mechanics}
\label{sec:warmup}

To explain our variational technique and its limitations, let us first consider the case of quantum mechanics with a non-trivial potential. We choose to study the pure quartic oscillator potential for which the path integral representation of the  quantum mechanical partition function at finite temperature $T=\frac{1}{\beta}$ is given by
\be
Z={\rm Tr}e^{-\beta {\cal H}}=\int {\cal D}\phi e^{-\int_0^\beta d\tau\left[\frac{1}{2}\partial_\tau \phi \partial_\tau \phi+\lambda \phi^4\right]}\,,
\ee
with $\phi(\tau)$ obeying periodic boundary conditions on the interval $\tau\in[0,\beta]$. Here $\lambda$ is the coupling constant of the theory, which in our units has mass dimension cubed. In quantum mechanics, inserting a complete set of energy eigenstates $|n\rangle$ with eigenvalues $E_n$, it is possible to rewrite the partition function as
\be
\label{Zdef1}
Z=\sum_{n=0}^\infty \langle n | e^{-\beta {\cal H}}|n\rangle = \sum_{n=0}^\infty e^{-\beta E_n}\,.
\ee
Because $\lambda$ is the only dimensionful parameter of the theory at zero temperature, simple dimensional analysis leads to scaling of energy eigenvalues with the coupling as
\be
E_n=\lambda^{\frac{1}{3}}\hat E_n\,,
\ee
where $\hat E_n$ are pure numbers. In particular, this leads to the behavior of the partition function at low temperature $\beta\rightarrow \infty$ as
\be
Z=\sum_{n=0}^\infty e^{-\beta \lambda^{\frac{1}{3}}\hat E_n}\simeq e^{-\beta \lambda^{\frac{1}{3}}\hat E_0}\,,
\ee
with
\be
\label{E00}
\hat E_0\simeq 0.667986\ldots\,,
\ee
the ground-state energy for the pure quartic oscillator\cite{Bender:1977dr,Hioe:1978jj,Romatschke:2020qfr}.

The path integral may be discretized on a lattice with $N$ sites by replacing the differentials in the action with finite differences,
\be
\partial_\tau \phi(\tau)=\frac{\phi(\tau+a)-\phi(\tau)}{a}\,,
\ee
where $a=\frac{\beta}{N}$ is the lattice spacing. The discretized path integral for quantum mechanics becomes \cite{Laine:2016hma}
\be
Z_N=\int \prod_{k=0}^{N-1} \frac{d \phi_k}{\sqrt{2\pi a}} e^{-a \sum_{k=0}^{N-1}\left[\frac{\left(\phi_{k+1}-\phi_k\right)^2}{2a^2}+\lambda \phi_k^4\right]}\,,
\ee
where $\phi_k\equiv \phi(\tau=k a)$ and periodic boundary conditions imply $\phi_{N}=\phi_0$.

However, it is also possible to discretize the path integral in a different basis than position basis. In fact, one can use any set of complete basis vectors to discretize the partition function. Denoting such a complete set of basis vectors by $|n\rangle$, with $n=0,1,\ldots$ we have
\be
Z=\lim_{N\rightarrow \infty}\sum_{n_1,n_2,\ldots,n_N}\langle n_1 | e^{-\frac{\beta {\cal H}}{N}}|n_2\rangle \langle n_2 | e^{-\frac{\beta {\cal H}}{N}}|n_3\rangle\ldots \langle n_N | e^{-\frac{\beta {\cal H}}{N}}|n_1\rangle\,,
\ee
where the periodicity of the trace enforced $n_N=n_1$. Considering the individual factors as ``transfer'' matrices
\be
T_{nm}\equiv \langle n|e^{-\frac{\beta {\cal H}}{N}}|m\rangle\,,
\ee
one observes that the partition function can be written as
\be
Z=\lim_{N\rightarrow \infty} {\rm Tr}\, T^{N}\,,
\ee
which is the familiar transfer-matrix form of the partition function. In the continuum limit $N\rightarrow \infty$, we have
\be
T_{nm}\simeq \delta_{nm}-\frac{\beta}{N}\langle n|{\cal H}|m\rangle+{\cal O}\left(\frac{\beta^2}{N^2}\right)\,.
\ee

 The transfer matrix entries can be calculated as 
  \be
  \langle n|{\cal H}|m\rangle=\int dx \langle n|x\rangle \langle x|{\cal H}|m\rangle=\int dx \psi_n(x) {\cal H} \psi_m(x)\,,
  \ee
  where in position basis the Hamiltonian operator becomes
  \be
  \label{theHamiltonian}
  {\cal H}=-\frac{1}{2}\frac{d^2}{dx^2}+\lambda x^4\,.
  \ee
  
  For numerical purposes, it is convenient to truncate the set of eigenvectors $|n\rangle$ at some level $n=K$. Denoting the ordered eigenvalues of the rank $K$ matrix $\langle n | {\cal H}|m\rangle$ by $e_0,e_1,\ldots e_{K-1}$, we have
  \be
  \label{zzz}
Z=\lim_{K\rightarrow \infty}\sum_{n=0}^{K-1} e^{-\beta e_n}\,,
\ee
or equivalently
\be
\lim_{K\rightarrow \infty} e_n=E_n\,,
\ee
with $E_n$ the ordered eigenvalues of the Hamiltonian ${\cal H}$.

Let us now see how this works in practice by using the scaled Hermite functions as the basis set for the transfer matrix:
\be
\langle x|n\rangle = \psi_n(x)=\frac{\omega^{\frac{1}{4}} e^{-\frac{\omega x^2}{4}}H_n(\sqrt{\omega} x)}{\sqrt{2^n n! \sqrt{\pi}}}\,,
  \ee
  where $\omega$ at this point is a free parameter. The basis functions obey
  \be
  \left[-\frac{1}{2}\frac{d^2}{dx^2}+\frac{\omega^2 x^2}{2}\right]\psi_n(x)=\frac{\omega (2n+1)}{2}\psi_n(x)\,,
  \ee
  so that 
  \be
  \langle n|{\cal H}|m\rangle=\int dx \psi_n(x)\left[\frac{\omega (2m+1)}{2}-\frac{\omega^2 x^2}{2}+\lambda x^4\right]\psi_m(x)\,.
  \ee

  The matrix elements are easily evaluated using the properties of the Hermite functions. A further simplification can be obtained by recalling that the eigenvalues of the Hamiltonian separate into parity-even and parity-odd sectors, which are spanned by the even-ordered (odd-ordered) Hermite functions, respectively. Since the parity-even and parity-odd sectors do not mix, we can consider the transfer matrix for these sectors separately.  As a specific example, for $K=2$ we take $\psi_0,\psi_2$ to span the parity-even sector and hence we get
  \be
  \langle n|{\cal H}|m\rangle=\left(\begin{array}{cc}
    \frac{3 \lambda}{4\omega^2}+\frac{\omega}{4} &
    \frac{1}{\sqrt{2}}\left(\frac{3\lambda}{\omega^2}-\frac{\omega}{2}\right)\\
      \frac{1}{\sqrt{2}}\left(\frac{3\lambda}{\omega^2}-\frac{\omega}{2}\right) &\frac{39 \lambda}{4\omega^2}+\frac{5\omega}{4}
   
    \end{array}\right)\,,
  \ee
 with the eigenvalues given by
  \be
  e_{0,2}=\frac{1}{4\hat{\omega}^2}\left(21+3\hat{\omega}^3\pm\sqrt{396+48\hat{\omega}^3+6 \hat{\omega}^6}\right)\lambda^{\frac{1}{3}}\,,\quad \hat{\omega}=\frac{\omega}{\lambda^{\frac{1}{3}}}\,.
  \ee

  Since $\omega$ is an arbitrary parameter, it should cancel out for the evaluation of the eigenspectrum $E_n$ of the Hamiltonian ${\cal H}$. However, since we have truncated the basis set, $\omega$ may appear in the approximate eigenvalues $e_n$. The dependence of $e_n$ on $\omega$ is expected to become weaker and weaker as the truncation of the basis set is removed: $K\rightarrow \infty$. For finite truncation $K$, a useful approximation is found by taking $\omega$ to be the point of minimal variation of $e_0$, or simply as
  \be
  \omega:\quad {\rm min}(e_0)\,.
  \ee
  For the case at hand, this results in
  \be
  \hat e_{0,2}=\frac{e_{0,2}}{\lambda^{\frac{1}{3}}}\simeq \left\{0.668668,4.87173\right\}\,,
  \ee
  where one should remark that even for $K=2$ the value of $\hat e_0$ is a good approximation for the ground state energy $\hat E_0$ given in (\ref{E00}). Increasing $K$ and using only the parity-even basis functions then leads to the following results for the ground state energy and relative error with respect to (\ref{E00}):

  \begin{centering}
    
    \hfill
    \begin{tabular}{|c|cccc|}
    \hline
    K & 2 & 3 & 4 & 5\\
    \hline
    $\hat{\lambda}_0$ & 0.668668 & 0.668041 & 0.667992 & 0.667987 \\
    \hline
    rel. err & $10^{-3}$ & $10^{-4}$ & $10^{-5}$ & $10^{-6}$\\
    \hline
    \end{tabular}
    \hfill
  \end{centering}
 
 \vspace*{1cm}
  As can be seen from this table, the variational method is quite efficient for obtaining the ground-state energy of the system.

\subsection{Phase transition point for the critical $\phi^4$ theory in 2d}
\label{sec:2dphipos}

To study the critical point in scalar $\phi^4$ theory in two dimensions defined by the classical Euclidean action (\ref{latticeaction}), we scale $\phi\rightarrow \frac{\phi}{\sqrt{a}}$. Subsequently changing notation as
\be
\phi(\tau,x_k)=\phi_k(\tau)\equiv x_k
\ee
leads to
\be
S_{(D)}=\int_0^\beta d\tau \sum_{k=0}^{D-1} \left[\frac{\dot{x}^2_k}{2}+\frac{\left(x_{k+1}-x_k\right)^2}{2a^2}+\frac{m_B^2 x^2_k}{2}+\frac{\lambda}{a} x^4_k\right]\,,
\ee
which is the action for quantum mechanics in $D$ dimensions with potential
\be
\label{qmpot}
V(\vec{x})=\frac{m_B^2 \vec{x}^2}{2}+\sum_{k=0}^{D-1}\left[\frac{(x_{k+1}-x_k)^2}{2a^2}+\frac{\lambda}{a}x_k^4\right]\,,
\ee
for the Hamiltonian ${\cal H}=\frac{\vec{p}^2}{2}+V(\vec{x})$.

As a consequence, following the same steps as for the quantum mechanics problem in the previous section, the partition function for 2d scalar quantum field theory can be written as
\be
\label{Zrecast}
Z={\rm Tr}e^{-\beta {\cal H}}=\lim_{K\rightarrow \infty}\sum_{n=0}^{K-1}e^{-\beta e_n}\,,
\ee
where $e_n$ are the eigenvalues of the truncated matrix elements $\langle n|{\cal H}|m\rangle$. One can once more use the property of the Hermite functions to form a complete and orthogonal basis set to construct the states $|n\rangle$. For instance, for the parity-even and parity-odd sector of the theory, respectively, one can consider the states
\be
\langle \vec{x} |0\rangle = \prod_{k=0}^{D-1}\psi_0(x_k), \quad \langle \vec{x} |1\rangle=\frac{1}{\sqrt{D}}\left[\psi_1(x_0)\prod_{k=1}^{D-1}\psi_0(x_k)+{\rm perm}.\right]\,.
\ee
Note that these states will in general \textit{not} correspond to the lowest-energy eigenvectors of the Hamiltonian, but they are reasonably representative examples of parity-even and parity-odd states. Using these states, one has for the matrix elements of the Hamiltonian
\ba
\langle 0 |{\cal H}|0\rangle&=&D\left[\frac{ \omega}{4}+\frac{m_B^2+2 a^{-2}}{4\omega} +\frac{3\lambda}{4 a \omega^2}\right]\,,\quad\\
\langle 1 |{\cal H}|1\rangle&=&\frac{\omega(D+2)}{4}+\frac{(D+2)m_B^2}{4\omega}+\frac{D}{2a^2\omega} +\frac{3(D+4)\lambda}{4 a \omega^2}\,,\nonumber\\
\langle 1 |{\cal H}|0\rangle&=&0\,.
\ea

In the crudest variational approximation, we truncate both the parity-even and parity-odd sector at this order, such that $e_0,e_1$ are given by the matrix elements $\langle 0 |{\cal H}|0\rangle,\langle 1 |{\cal H}|1\rangle$ from above. For $m_B^2>0$, we always have $e_1>e_0$, indicating that the ground state of the theory is parity-symmetric. However, for $m_B^2<0$ the ordering of $e_0,e_1$ may reverse, and one may find $e_1<e_0$, which indicates that the ground state of the system is in a parity-odd state, or a broken phase. In the continuum limit of 2d $\phi^4$ theory, one does not expect $e_1<e_0$, but instead $e_1=e_0$ for $m_B^2<m_{B,{\rm crit}}^2$, indicating the presence of a second order phase transition at $m_B^2=m_{B,{\rm crit}}^2$ (cf. \cite[Eq.~(101)]{onsager1944crystal}). In our variational method for finite system size $D$, the value of $m_{B,{\rm crit}}^2$ where the ordering of $e_1,e_0$ changes corresponds instead to a \textit{first} order phase transition. This can be seen by considering the partition function representations (\ref{Zdef1}) and $Z=\int {\cal D}\phi e^{-S}$ and equating
\be
-\frac{1}{\beta L} \frac{\partial \ln Z}{\partial m_B^2}=\frac{1}{L}\frac{e^{-\beta E_0}\frac{\partial E_0}{\partial m_B^2}+e^{-\beta E_1}\frac{\partial E_1}{\partial m_B^2}+\ldots}{e^{-\beta E_0}+e^{-\beta E_1}+\ldots}=\frac{1}{2\beta L}\int d^2x \langle \phi^2(x)\rangle \,,
\ee
where $\langle {\cal O}\rangle\equiv \frac{1}{Z}\int {\cal D}\phi e^{-S}\cal O$. In the zero temperature limit $\beta\rightarrow \infty$, this quantity is continuous as a function of $m_B^2$, except at the point $m_B^2=m_{B,{\rm crit}}^2$ where the quantity jumps from the first term for $m_B^2>m_{B,{\rm crit}}^2$ and $E_1>E_0$ discontinuously to the second term for $m_B^2<m_{B,{\rm crit}}^2$ and $E_1<E_0$. This is because while $E_0=E_1$ at $m_B^2=m_{B,{\rm crit}}^2$, the derivatives differ: $\frac{\partial E_0}{\partial m_B^2}\neq \frac{\partial E_1}{\partial m_B^2}$.
This implies that one can use the energy level difference $\Delta\equiv e_1-e_0$ between parity-odd and parity-even sectors as an order parameter for the first order transition for any finite $D$. This is particularly welcome for the variational approach since this approach is most economical for reporting the energy spectrum of the theory, as indicated in section \ref{sec:warmup}.

In the limit of $D\rightarrow \infty$, we expect the line of first order phase transitions to end in a second-order critical point where the energy levels become degenerate $E_1=E_0$ for all $m_B^2<m_{B,{\rm crit}}^2$. The second order derivative then leads to
\be
\frac{1}{\beta L} \frac{\partial^2 \ln Z}{\partial^2 m_B^2}=\frac{1}{4\beta V}\int_{x,y} \left[\langle \phi^2(x)\phi^2(y)\rangle-\langle\phi^2(x)\rangle \langle \phi^2(y)\rangle\right]=\frac{1}{2}\int d^2x\langle \phi(x)\phi(0)\rangle^2\,. 
\ee
For a free boson of mass $M$, we have  $\langle \phi(x)\phi(0)\rangle=\int \frac{d^2k}{(2\pi)^2}\frac{e^{i k\cdot x}}{k^2+M^2}=\frac{K_0(M\sqrt{x^2})}{2\pi}$ with $K_0$ denoting the modified Bessel function of the second kind. For large $\sqrt{x^2}$, one has
\be
K_0(M \sqrt{x^2})\rightarrow e^{-M \sqrt{x^2}}\sqrt{\frac{\pi}{2 M \sqrt{x^2}}}\,,
\ee
which identifies $M$ as the inverse \textbf{correlation length} $C_L=\frac{1}{M}$ of the theory. One then finds
\be
\label{corrdef}
\frac{1}{\beta L} \frac{\partial^2 \ln Z}{\partial^2 m_B^2}=\frac{1}{2}\int \frac{d^2 x}{(2\pi)^2}K_0^2(M\sqrt{x^2})=\frac{1}{8 \pi M^2}=\frac{C_L^2}{8\pi}\,,
\ee
which implies that $C_L$ diverges when $M\rightarrow 0$ or when the second derivative of the partition function diverges. This is the case for a second-order phase transition.

Using the matrix elements from above to estimate $e_0,e_1$ we have
\be
\label{k2result}
K=2:\quad e_1-e_0=\frac{6 \lambda+a \omega (m_B^2+\omega^2)}{2 a \omega^2}\,,
\ee
so that we expect a transition from symmetric to broken phase for $\Delta=0$ at
\be
\label{k1cir}
K=2: \quad m_{B,{\rm crit}}^2=-\omega^2-\frac{6 \lambda}{a \omega}\,.
\ee

The parameter $\omega$ is once again fixed by the requirement ${\rm min}(e_0)$ which together with (\ref{k1cir}) leads to
\be
K=2: \quad \omega=\frac{1}{a}\,,
\ee
so that
\be
K=2: \quad \frac{m_{B,{\rm crit}}^2}{\lambda}=-\frac{1}{a^2\lambda}-6 \,.
\ee

\subsection{Numerical evaluations for a higher number of basis states}
\label{numerical}

It is also possible to increase the number of basis states for fixed D by evaluating matrix elements numerically. To this end, let us define parity-even and parity-odd basis states as
\be
\langle \vec{x}|n_1 n_2 n_3\ldots n_D\rangle=\psi_{n_1}(x_0)\psi_{n_2}(x_1)\ldots \psi_{n_D}(x_{D-1})\,,
\ee
where for the parity-even states we require $\sum_{i=1}^D n_i={\rm even}$, while 
for the parity-odd states we require $\sum_{i=1}^D n_i={\rm odd}$. We then order the states according to their naive harmonic oscillator energy values, e.g.
$E_{\rm triv}(n_1,n_2,\ldots,n_D)=\frac{D}{2}+\sum_{i=1}^D n_i$. For each pair of states $|n_1 n_2 n_3\ldots n_D\rangle, |m_1 m_2 m_3\ldots m_D\rangle$ we calculate the matrix element as
\be
\langle m_1 m_2\ldots m_D |{\cal H} |n_1 n_2 n_3\ldots n_D\rangle
={\cal H}_{n_1 n_2 \ldots n_D}^{m_1 m_2 \ldots m_D}\,,
\ee
where 
\ba
{\cal H}_{n_1 n_2 \ldots n_D}^{m_1 m_2 \ldots m_D}&=&\sum_{i=1}^D \langle m_i | \frac{p^2}{2}+\frac{x^2}{2}\left(m_B^2+\frac{2}{a^2}\right)
+\frac{\lambda x^4}{a}|n_i\rangle\prod_{j=1,j\neq i}^D \delta_{m_j n_j}
\nonumber\\
&&
-\frac{1}{a^2}\sum_{i=1}^D\langle m_i|x|n_i\rangle \langle m_{i+1}|x|n_{i+1}\rangle
\prod_{j=1,j\neq i,j\neq i+1}^D \delta_{m_j n_j}\,.
\ea
Note that the matrix elements in ${\cal H}_{n_1 n_2 \ldots n_D}^{m_1 m_2 \ldots m_D}$ are quantum-mechanical matrix elements which can be evaluated as
\be
\langle m| \frac{p^2}{2}+\frac{x^2}{2}\left(m_B^2+\frac{2}{a^2}\right)+\frac{\lambda}{a} x^4|n\rangle=\left\{\begin{array}{c}\frac{\lambda \sqrt{(n+1)(n+2)(n+3)(n+4)}}{4 a \omega^2}\,,\ m=n+4\\
\frac{\sqrt{(n+1)(n+2)}}{2 \omega}\left[\frac{\left(m_B^2+\frac{2}{a^2}-\omega^2\right)}{2} +\frac{\lambda (3+2n)}{a\omega}\right]\,,\  m=n+2\,,\\
\left(n+\frac{1}{2}\right)\left(\frac{\omega}{2}+\frac{m_B^2+\frac{2}{a^2}}{2\omega}\right)+\frac{3\lambda (1+2n+2n^2)}{4 a\omega^2}\,,\ n=m\,,\\
0\,,\ {\rm else}\,,
\end{array}\right.
\ee
where \cite[Eq.~(18.17.25)]{NIST:DLMF} has been used to evaluate the integrals of Hermit functions with weights $x^2,x^4$, respectively. Similarly, using \cite[Eq.~(18.17.26)]{NIST:DLMF} one finds
\be
\langle m|x|n \rangle=\left\{\begin{array}{c}
\sqrt{\frac{n+1}{2\omega}}\,,\ n=m+1\,,\\
0\,, {\rm else}\,.
\end{array}
\right.
\ee

\begin{figure}[t]
  \includegraphics[width=.8\linewidth]{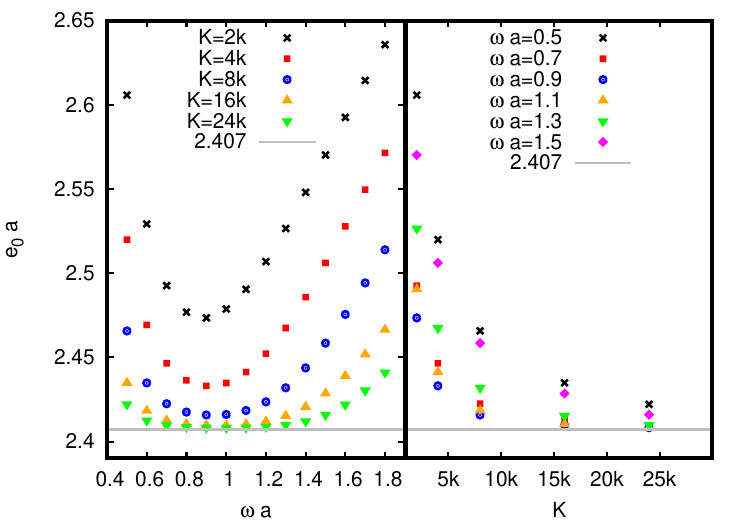}
  \caption{\label{fig:zero} Dependence of $e_0 a$ for $D=5$, $m_B^2 a^2=-0.5$ and $\lambda a^2=0.05$ on variational parameter $\omega a$ and number of basis states $K$. Left: results as a function of the variational parameter $\omega a$ for various values of $K=2k,4k,8k,16k$ and $K=24k$, indicating the emergence of a minimum and plateau around $\omega a\simeq 1$. Right: results as a function of the number of basis states $K$ for values of $\omega a=0.5,0.7,0.9,1.1,1.3$ and $\omega a=1.5$, indicating minimal sensitivity to $K$ for $\omega a\simeq 1$. Both methods point to the same result $e_0 a\simeq 2.407$.}
  \end{figure}

\section{Comparing Variational and Saddle-Point Methods}

We fix units by expressing all quantities in terms of the ``spatial'' lattice spacing $a$. For the variational method, fixing the number of spatial sites $D$ and the dimensionless coupling $\lambda a^2$, using the matrix elements, we vary $\omega a$ numerically to find the minimum values of $e_1,e_0$ for any given value of $m_B^2 a^2$. Representative results from this procedure for $D=5$, $m_B^2 a^2=-0.5$ and $\lambda a^2=0.05$ are shown in Fig.~\ref{fig:zero}. It can be seen that a minimum in $\omega a$ develops and becomes broader as the number of basis states is increased. Fig.~\ref{fig:zero} also demonstrates that alternatively fixing $\omega a$ and finding the minimal variation with respect to the number of basis states leads to the same result.

\begin{figure}[t]
  \includegraphics[width=.48\linewidth]{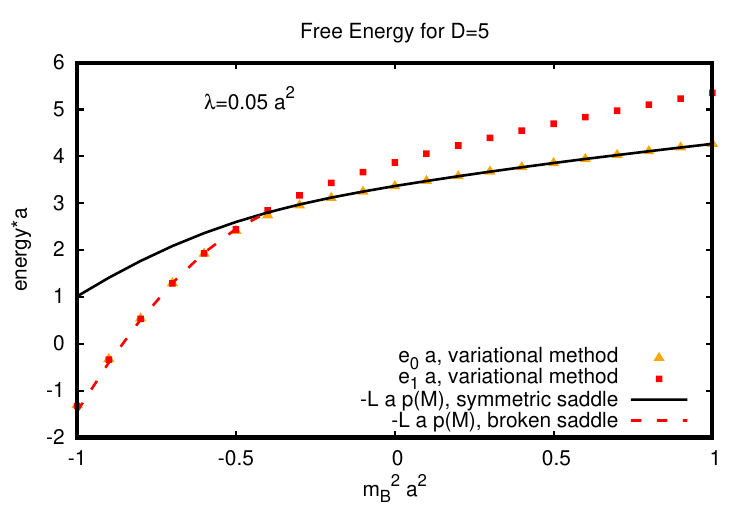}
  \hfill
   \includegraphics[width=.48\linewidth]{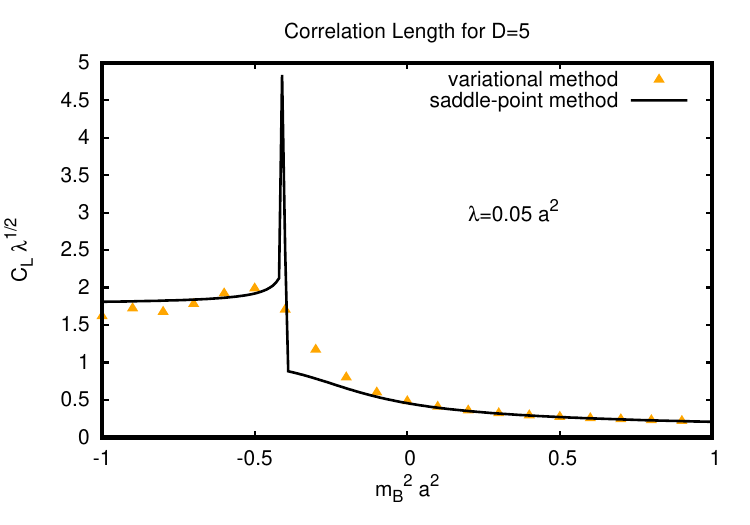}
   \caption{\label{fig:one} Left: Dependence of energies on $m_B^2 a^2$ for $D=5$, $\lambda a^2=0.05$. Shown are results for $e_0 a,e_1 a$ for $K=16k$ at the minimum of the variational parameter $\omega a$ (symbols). Also shown are results for the free energy from the analytic saddle expansions in the symmetric and broken phase saddles Eq.~(\ref{saddlepoints}).
     Right: Correlation length $C_L$ defined in (\ref{corrdef}) as a function of $m_B^2 a^2$ for $D=5$, $\lambda a^2=0.05$. Shown are results for $K=16k$ at the minimum of the variational parameter $\omega a$ (symbols) and analytic saddle expansion  (full line).}
  \end{figure}

The dependence of $e_0 a,e_1 a$ as a function of the bare mass parameter $m_B^2 a^2$ is shown in Fig.\ref{fig:one} for the case of D=5, $\lambda a^2=0.05$. The values for $e_0, e_1$ were obtained by locating the minimum in the variational parameter $\omega a$ for $K=16k$ basis states. The results for the energies are compared to the free energies $-L a\, p(M),-L a \, \tilde p(\tilde M)$ obtained in the analytic saddle point expansion given in Eq.~(\ref{saddlepoints}) above. As can be seen, the saddle-point results for the symmetric saddle agrees quantitatively well with $e_0$ for $m_B^2\gtrsim -0.4$, whereas the saddle-point result for the broken saddle agrees quantitatively well with both $e_0,e_1$ for $m_B^2\lesssim -0.4$. This suggests that the broken phase saddle describes a phase consisting of superposition of parity-even and parity-odd states.

Also shown in Fig.~(\ref{fig:one}) is the dependence of the correlation length $C_L$ from (\ref{corrdef}) for the case of D=5, $\lambda a^2=0.05$ as a function of $m_B^2 a^2$. As can be seen, the correlation length in lattice units is small for $m_B^2 a^2\gtrsim 0$, but peaks around $m_B^2 a^2\simeq -0.5$. The results from the variational method for the correlation length is consistent with the analytic saddle-point approximation for these parameters, although a quantitative difference in the peak location for $C_L$ on the order of 25 percent can be observed in Fig.~(\ref{fig:one}).

Defining $\Delta=e_1-e_0$, the phase transition from symmetric to broken phase occurs when $\Delta=0$. For numerical stability in the variational method, however, it is advantageous to instead define the transition point as the location where the relative error $\Delta/e_0$ becomes smaller than one percent. Using this definition, we then lower $m_B^2 a^2$ from 0 until $\Delta/e_0<0.01$, which defines $m^2_{B,{\rm crit}}a^2$ for a fixed lattice and value of $\lambda a^2$. As examples, we find the following results for D=5 from the variational method, compared to the saddle-point method:

\begin{center}
\begin{tabular}{|c|c|c|c|c|c|}
  \hline
  & K=2k & K=4k & K=8k & K=16k & saddle-point method\\
\hline
  $\lambda a^2$ &  $m^2_{B,{\rm crit}} a^2$ &$m^2_{B,{\rm crit}} a^2$ & $m^2_{B,{\rm crit}} a^2$ & $m^2_{B,{\rm crit}} a^2$& $m^2_{B,{\rm crit}} a^2$  \\
  \hline
  0.2 & -1.58 & -2.0 & -2.1 & -1.954 & -1.22\\
  0.1 & -0.89 & -0.94 & -0.91 & -0.903 & -0.706\\
  0.05 & -0.52 & -0.67 & -0.52 & -0.514 & -0.408 \\
  0.01 &       &       &  -0.146   & -0.141 & -0.118\\
\hline
  \end{tabular}
  \end{center}

For the variational method, one observes that values of $m^2_{B,{\rm crit}} a^2$ for fixed $D,\lambda a^2$ stabilize as $K$ is increased. However, the value for $K$ where the method stabilizes increases as $D$ is increased, so that the variational method quickly becomes impractical for values of $D\gg 5$.
The value obtained for $m^2_{B,{\rm crit}} a^2$ in the variational method differs quantitatively from the saddle point method, but the relative discrepancy decreases as $\lambda a^2$ is decreased.

\begin{figure}[t]
  \includegraphics[width=.8\linewidth]{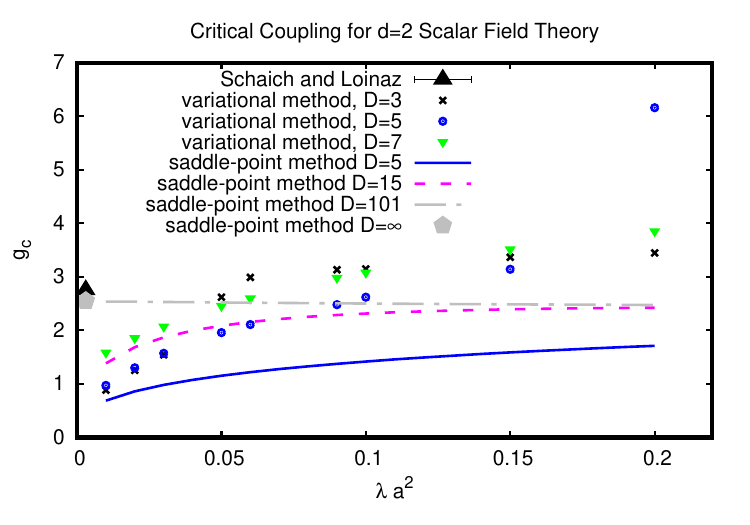}
  \caption{\label{fig:two} Critical coupling $g_c$ from the variational approach for various values of $D$ and $\lambda a^2$. For comparison, the continuum extrapolated result from lattice field theory studies \cite{Schaich:2009jk} as well as the continuum result from the analytic saddle expansion \cite{Romatschke:2026tam} are also indicated.}
  \end{figure}

Once $m^2_{B,{\rm crit}}a^2$ is determined, $m_{\rm crit}^2$ is calculated by numerically solving (\ref{gap}) using (\ref{gfreedef}) for a lattice with an odd number of transverse sites $D$,
\be
m_{\rm crit}^2 a^2 = m_{B,{\rm crit}}^2 a^2+6 \lambda a^2\left(\frac{1}{m_{\rm crit} a D}+2 \sum_{n=1}^{(D-1)/2}\frac{1}{\sqrt{\Omega_n^2a^2 D^2+m_{\rm crit}^2 a^2 D^2}}\right)\,,
\ee
which in turn lets us report a value for the critical coupling $g_c=\frac{ \lambda a^2}{m_{\rm crit}^2 a^2}$ for specific values of $D$.
%
Results for $g_c$ for various values of $D,\lambda a^2$ are shown in Fig.~\ref{fig:two}, along with the continuum-extrapolated result (\ref{gvalue}) from Ref.~\cite{Schaich:2009jk}, and the continuum result for the saddle-point method from Ref.~\cite{Romatschke:2026tam}, and the saddle-point method for $D=5,15,101$.

As can be seen from this comparison, the results from the variational method are qualitatively consistent with results from the saddle-point method. The saddle-point method indicates that finite transverse sizes of $D>15$ are necessary in order to observe the approach to the continuum limit for couplings $\lambda a^2\lesssim 0.2$. Since the largest transverse extent for the variational method studied in this work was $D=7$, we do not expect -- nor do we observe -- quantitative agreement of the variational results and the continuum extrapolations from lattice field theory \cite{Schaich:2009jk} or newer lattice Monte Carlo studies finding $g_c=2.7637(35)$ \cite{Bronzin:2018tqz}. Note that other approaches are broadly in line with the lattice results, for instance perturbative methods ($g_c=2.807(34)$ \cite{Serone:2018gjo}, $g_c=2.779$ \cite{Heymans:2021rqo}), tensor networks ($g_c=2.728(14)$ \cite{Kadoh:2018tis}), Hamiltonian truncation ($g_c=2.76(3)$ \cite{Elias-Miro:2017tup}) and matrix product states ($g_c\simeq 2.765$ \cite{Vanhecke:2019pez}).

The variational results for fixed $D$ are consistently above the saddle-point method. This agrees with the $D\rightarrow \infty$ result from the saddle-point method, already given in Ref.~\cite{Romatschke:2026tam} as
\be
g_c\simeq 2.5527\,,
\ee
which is approximately 6 percent below the result (\ref{gvalue}), cf. Fig.~\ref{fig:two}. However, it should be noted that obtaining precise quantitative results using our variational method is prohibitively computationally expensive because $K$ must be increased significantly as the number of transverse fields $D$ is increased. For values of $K,D$ that are easily accessible using desktop computing resources, only qualitative results for the continuum theory critical coupling $g_c$ can be extracted from the variational approach. On the other hand, the result from the saddle-point method for $g_c$ in the continuum limit is qualitatively consistent with other approaches in the continuum limit, but differs quantitatively on the ten-percent level.

\section{Summary and Conclusions}

In this work, we introduced a variational method to study scalar $\phi^4$ theory in two Euclidean dimensions in lattice discretization. The method was found to be numerically reliable for lattices with small transverse sizes, whereas the longitudinal direction can be treated as a continuum. Using the variational method, we were able to identify a critical line in the plane of bare lattice parameters corresponding to the breaking of parity symmetry. We gave arguments that this critical line ends in the critical point of the two-dimensional scalar field theory in the continuum limit.

For the same lattice configurations as the variational approach, we evaluated self-duality relations using the saddle-point expansions from Ref.~\cite{Romatschke:2026tam} and compared to the results from the variational approach. We found that the self-duality relations are qualitatively reliable for the overall phase structure of the theory, and that observables such as the free energy are in quantitative agreement with the variational method. Quantitative differences on the 10 to 25 percent level were observed for more refined observables, such as the peak location of the correlation length, which involves a second-order derivative of the free energy.

While our numerical implementation of our variational method did not allow continuum extrapolation of variational results, the continuum extrapolation of the saddle-point expansion was straightforward and was compared to the continuum-extrapolated results for the critical coupling from other methods \cite{Schaich:2009jk}, finding a quantitative difference of approximately six percent.

In conclusion, the self-duality relations as implemented through the saddle point expansions in Ref. ~\cite{Romatschke:2026tam} were found to be qualitatively reliable for the phase diagram of scalar field theory in two dimensions. They were found to be quantitatively reliable for quantities such as the free energy, but differences on the 10 to 25 percent level were encountered for derivative quantities such as the correlation length.

Overall, the agreement between analytic method and numerical results is sufficiently good to encourage application to other systems, in particular the phase diagram for scalar field theory in three and four dimensions. We leave these applications to future studies.

  \section{Acknowledgments}

  This work was partially supported by the Department of Energy, DOE award No DE-SC0017905. This material is based upon work supported by the NSF National Center for Atmospheric Research, which is a major facility sponsored by the U.S. National Science Foundation under Cooperative Agreement 1852977.
 PR would like to thank Scott Lawrence for multiple discussions on this topic, and the Nuclear Theory and High Energy Physics Groups at Los Alamos National Laboratory for their hospitality.

\bibliography{enormous}
\end{document}